\documentclass[aps,twocolumn,showpacs]{revtex4}
\usepackage{amsfonts}
\usepackage{amsmath}
\usepackage{amssymb}
\usepackage{graphicx}
\usepackage{fancyheadings}
\usepackage{times,xspace,color}
\usepackage{bm}
\usepackage{multirow}

\def\fS{{\sf{T}}}

\def\bbbc{{\mathchoice {\setbox0=\hbox{$\displaystyle\rm C$}\hbox{\hbox
to0pt{\kern0.4\wd0\vrule height0.9\ht0\hss}\box0}}
{\setbox0=\hbox{$\textstyle\rm C$}\hbox{\hbox
to0pt{\kern0.4\wd0\vrule height0.9\ht0\hss}\box0}}
{\setbox0=\hbox{$\scriptstyle\rm C$}\hbox{\hbox
to0pt{\kern0.4\wd0\vrule height0.9\ht0\hss}\box0}}
{\setbox0=\hbox{$\scriptscriptstyle\rm C$}\hbox{\hbox
to0pt{\kern0.4\wd0\vrule height0.9\ht0\hss}\box0}}}}

\newcommand{\bk}{\mathbf{k}}
\newcommand{\bQ}{\mathbf{Q}}

\preprint{}  
\newenvironment{Quote}{\begin{list}{$\bullet$}{\setlength{\leftmargin}{0.25cm}%
\setlength{\labelsep}{0.1cm}} \item[] } {\end{list}}

\begin{document}

\title{Integrable models for asymmetric Fermi superfluids: Emergence of
a new exotic pairing phase}
\author{J.~Dukelsky$^{1}$, G. Ortiz$^{2}$, S.M.A. Rombouts$^3$, and K.
Van Houcke$^3$}
\address{$^{1}$ Instituto de Estructura de la Materia, CSIC, Serrano
123, 28006 Madrid, Spain \\
$^{2}$ Theoretical Division, Los Alamos National Laboratory, Los Alamos,
New Mexico 87545, USA \\
$^{3}$ Universiteit Gent - UGent, Vakgroep Subatomaire en
Stralingsfysica, Proeftuinstraat 86, B-9000 Gent, Belgium}

\date{Received \today }

\begin{abstract}
We introduce an exactly-solvable model to study the competition between
the Larkin-Ovchinnikov-Fulde-Ferrell (LOFF) and breached-pair
superfluid in strongly interacting ultracold asymmetric Fermi gases.
One can thus investigate homogeneous and inhomogeneous states on an
equal footing and establish the quantum phase diagram. For certain
values of the filling and the interaction strength, the model exhibits
a new stable exotic pairing phase  which combines an inhomogeneous
state with an interior gap to pair-excitations.  It is proven that this
phase is the exact ground state in the strong coupling limit, while
numerical examples in finite lattices show that also at finite
interaction strength  it can have lower energy than the breached-pair
or LOFF states.
\end{abstract}

\author{}
\maketitle

Superconducting and Fermi superfluid phenomena have been a subject of
fascination since their discovery. Both phenomena are direct
macroscopic-scale manifestations of quantum physics, with the electric
charge of their relevant microscopic constituents being the crucial
factor differentiating them. Interest in  their various fundamental
aspects has increased recently because advances in the field of
ultracold  atomic Fermi gases \cite{Regal,becbcs} are leading to new
experimental probes to investigate unexplored territory, with
consequences in condensed matter as well as high energy physics (e.g.,
the physics in the core of neutron stars).

Of particular importance is the nature of their ground states (GSs)
under various external conditions since novel thermodynamic phases
might show up. The present manuscript studies the relative vacuum
stability of a two-species fermion gas as a function of the pairing
interaction strength $g$ and different species population. Without loss
of generality we denote the two species as $a$ and $b$ with densities
$\rho_{a(b)}$. Differences among the two species could be related to
their masses $m_{a(b)}$, spin or hyperfine states. Asymmetry, in
general, makes pairing less favorable and questions about  the nature
of the resulting competing phases might arise.  To this end, we
introduce a model system that will prove to be exactly solvable and
which displays the competition between the
Larkin-Ovchinnikov-Fulde-Ferrell (LOFF) \cite{loff}, breached-pair (or
Sarma) \cite{sarma,liu}, deformed Fermi-surface superfluid
\cite{sedra}, and segregated phases \cite{bedaque}. Its quantum phase
diagram as a function of the asymmetry density $\delta
\rho=(\rho_a-\rho_b)/(\rho_a+\rho_b)$ and the coupling strength $g$ has
been recently studied, at the mean-field level, in the one-channel
\cite{phase} and two-channel models \cite{phase2}. In our exact
solution, albeit in a finite-size lattice, we find different regimes of
stability for these various phases. A key result is the prediction of a
new exotic inhomogeneous phase characterized by a particular
center-of-mass momentum of the condensed pairs.
This phase is the exact GS in the large-$g$ limit.

Consider $N_a$ and $N_b$ fermionic atoms confined to a $D$-dimensional
box of volume $V$, i.e. $\rho_{a(b)}=N_{a(b)}/V$, with periodic
boundary conditions and $g<0$. (The exact solvability of the problem is
not restricted to these latter conditions but for notational
convenience we will specialize to this case).
The following model Hamiltonian contains the right ingredients to
study the competition between the various phases
\begin{equation}
H=\sum_{\bk}( \varepsilon_{\bk}^{a} \ n^a_{\bk}+ \varepsilon_{\bk}^{b} \
n^b_{\bk})+2g\sum_{\bk,\bk^{\prime}}a_{\bk+\bQ}^{\dagger}
b_{-\bk}^{\dagger} b_{-\bk^{\prime}}^{\;}a_{\bk^{\prime}+\bQ}^{\;} ,
\label{hamil}
\end{equation}
where $a_{\bk}^{\dagger}$($b_{\bk}^{\dagger}$) creates a particle
of type $a$($b$) with momentum $\bk$ and
$n^a_\bk=a_{\bk}^{\dagger}a^{\;}_{\bk}$,
$n^b_\bk=b_{\bk}^{\dagger}b^{\;}_{\bk}$.  The pairing interaction
scatters pairs with center-of-mass momentum $\bQ$ and ``band''
energies $\varepsilon_\bk^\alpha=\epsilon_{\bk}/2 m_\alpha$
($\alpha=a,b$), with $\epsilon_\bk$ representing an arbitrary
dispersion (including a non-rotational-invariant one
\cite{sedra}).

The quantum integrability and exact solvability of the Hamiltonian
(\ref{hamil}) can be derived using an $su(2)$ algebra
\begin{equation}
{\tau}_{\bk,\bQ}^{+}=a_{\bk+\bQ}^\dagger
b_{-\bk}^\dagger =(\tau _{\bk,\bQ}^{-})^{\dagger}\
,\
{\tau}_{\bk,\bQ}^{z}=\frac{1}{2}(n^a_{\bk+\bQ}+n^b_{-\bk}-1)\
, \label{set1}
\end{equation}
and a second, independent, realization of $su(2)$
\begin{equation}
{S}_{\bk,\bQ}^{+}=a_{\bk+\bQ}^\dagger
b_{-\bk}^{\;} =(S_{\bk,\bQ}^{-})^{\dagger}\ ,\
{S}_{\bk,\bQ}^{z}=\frac{1}{2}(n^a_{\bk+\bQ}-n^b_{-\bk})\
. \label{set2}
\end{equation}
These two mutually commuting algebras are often referred to as  charge
and spin $su(2)$ realizations, respectively.
Using the algebraic techniques of the Richardson-Gaudin model~\cite{NB},
one can write down a complete set of integrals of motion
$R^\fS_{\bk,\mathbf{Q}}$, with $[R^\fS_{\bk,\mathbf{Q}},
R^{\fS'}_{\bk',\mathbf{Q}}]=0$  (for $\fS,\fS'=\tau,S$):
$R_{\bk,\bQ}^{\fS}=\fS_{\bk,\bQ}^{z}+ 2g_\fS\sum_{\bk^{\prime}(\neq
\bk)} X_{\bk\bk^{\prime}}^\fS \ \vec{\fS}_{\bk,\bQ}\cdot
\vec{\fS}_{\bk^\prime,\bQ}$,
where $X_{\bk\bk^{\prime}}^\fS=1/(\eta^\fS_{\left( \bk,\bQ \right)}
-\eta^\fS_{\left( \bk^{\prime},\bQ\right)} )$,
with arbitrary functions $\eta^\fS$ depending upon $\bk$ and
$\bQ$, and $g_\fS$  are the coupling constants. Their complete set
of eigenvectors are of the form
\begin{eqnarray}
\left\vert \Psi \right\rangle =\prod\limits_{\ell=1}^{M_\fS}\left(
\sum_{\bk}\frac{1}{2\eta_{\left(\bk,\bQ\right)}^\fS-E_{\ell}^\fS} \
\fS_{\bk,\bQ}^{+}\right) \left\vert \nu^\fS \right\rangle ,
\end{eqnarray}
where $\left\vert \nu^\fS\right\rangle \equiv $
$\prod\limits_{\bk}\left\vert \nu_{\bk,\bQ}^\fS\right\rangle$
is a quasispin vacuum state
defined by $\fS_{\bk,\bQ}^{-}\left\vert \nu^\fS\right\rangle=0$, and
$\fS_{\bk,\bQ}^{z} \left\vert \nu^\fS\right\rangle =
d_{\bk,\bQ}^\fS\left\vert \nu^\fS\right\rangle$,
with $d_{\bk,\bQ}^\fS=(2\nu^\fS_{\bk,\bQ}-\Omega_{\bk,\bQ})/4$,
$\Omega_{\bk,\bQ}=2$, and the seniority quantum number
$\nu_{\bk,\bQ}^\fS=1,0$,  which for the $su(2)$ pair algebra
(\ref{set1}) counts the number of unpaired fermions.
The complex spectral parameters $E_{\ell}^\fS$ satisfy the set of
non-linear equations
\begin{eqnarray}
\frac{1}{4g_\fS}-\sum_{\bk}\frac{d_{\bk,\bQ}^\fS}{2\eta^\fS_{\left(
\bk,\bQ\right)}-E_{\ell}^\fS}+\sum_{m \left( \neq \ell \right)
}\frac{1}{E_{\ell}^\fS-E_{m}^\fS}=0 \ . \label{Rich}
\end{eqnarray}

To simplify matters, and because our goal is to show exact
solvability of (\ref{hamil}), we will only consider dynamics in
the charge space (i.e., $g_S=0$, $g_\tau=g$, dropping the label
$\fS$). The total number of atoms is $N=N_a+N_b=2M+\nu$, where $M$
is the number of atom pairs and $\nu$ the number of unpaired ones.
Consider now the linear combination $H_\tau=2\sum_\bk \eta_{\left(
\bk,\bQ\right)} R_{\bk,\mathbf{Q}}^{\tau}=\sum_{\bk}2\eta_{\left(
\bk,\bQ\right)} \tau_{\bk,\bQ}^{z}+2g \sum_{\bk,\bk^{\prime}}
\tau_{\bk,\bQ}^{+}\tau_{\bk',\bQ}^{-} +C$, where
$C=3g\sum_{\bk}d_{\bk,\bQ} +g\left( N-L\right)^{2}/2- g\left(
N-L\right)$. Comparing $H_\tau$ with (\ref{hamil}) we immediately
see that they differ in the kinetic term. Making use of the spin
$su(2)$ algebra by adding a term of the form $2\sum_{\bk}
\xi_{(\bk,\bQ)} S^z_{\bk,\bQ}$,
it leads to (up to an irrelevant constant)
\begin{eqnarray}
H &=&\sum_{\bk} [ (\eta_{\left( \bk,\mathbf{Q}\right) }+\xi_{\left(
\bk,\mathbf{Q}\right)} ) n^a_{\bk+\bQ}+( \eta_{\left(
\bk,\mathbf{Q}\right)}-\xi_{\left( \bk,\mathbf{Q}\right)} ) n^b_{-\bk} ]
\nonumber \\
&+&2g\sum_{\bk,\bk^{\prime}}\tau_{\bk,\mathbf{ Q}}^{+}
\tau_{\bk^{\prime},\mathbf{Q}}^{-} .
\label{hamilt}
\end{eqnarray}
Identifying
$\eta_{\left(\bk,\bQ\right)}= \frac{1}{2} [
\varepsilon^a_{\bk+\bQ}+\varepsilon^b_{-\bk} ]$, and
$\xi_{\left(\bk,\bQ\right)}= \frac{1}{2} [
\varepsilon^a_{\bk+\bQ}-\varepsilon^b_{-\bk} ]$,
we get the Hamiltonian of Eq. (\ref{hamil}), after constraining the
vectors  $\bk+\bQ$ and $\bk$ to be in the same set.
The eigenvalue $E$ corresponding to the solutions of Eqs. (\ref{Rich})
are given by
\begin{equation}
   E = \sum_\bk \left( \varepsilon^a_{\bk+\bQ} \nu^a_{\bk+\bQ}
                      +\varepsilon^b_{-\bk} \nu^b_{-\bk} \right)
          + \sum_{\ell} E_{\ell},
\end{equation}
where $\nu^\alpha_\bk$ denotes the number of unpaired  $\alpha$
particles in the state with momentum $\bk$.
The space dimensionality of the problem enters through the band
dispersion $\varepsilon_{\bk}^\alpha$, and the effective degeneracies
$d_{\bk,\bQ}$  in  the exact solution (\ref{Rich}). The latter are in
turn defined by $\eta_{\left(\bk,\bQ\right)}$. Assuming space-inversion
symmetry ($\varepsilon_{\bk}^\alpha=\varepsilon_{-\bk}^\alpha$), the
degeneracies $\Omega_{\bk,\bQ}$ count the number of states [$\bk ,
\bQ$] with the same value of $\eta_{\left(\bk , \bQ\right)}$.

\begin{figure}[htb]
\hspace*{-0.3cm}
\includegraphics[angle=0,width=8.0cm,scale=1.]{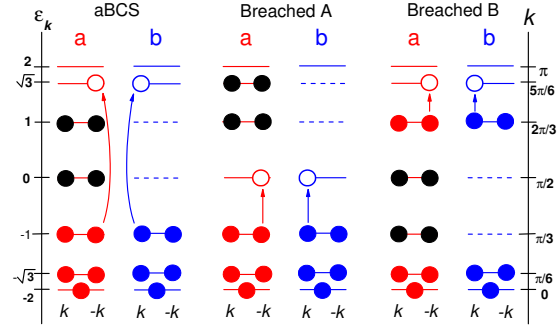}
\caption{Three $\bQ={\bf 0}$ configurations, asymmetric BCS (aBCS),
breached A, and breached B, as described in the text for a $D=1$
lattice with 12 sites, $N_a=9$ and $N_b=5$ atoms. The left vertical
axis displays the single-particle energies $\varepsilon_\bk^\alpha=-2
\cos \bk$ ($\alpha=a,b$) while the right one shows the corresponding
momenta $\bk$. An allowed pair-scattering process is indicated in each
case with arrows.
}
\label{fig1}
\end{figure}

\begin{figure}[htb]
\hspace*{-1.2cm}
\includegraphics[angle=0,width=7.5cm,scale=1.]{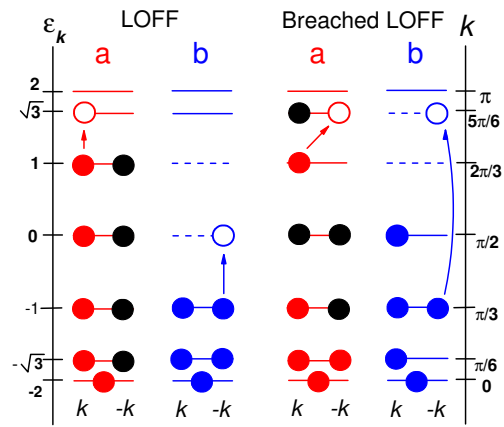}
\caption{Two possible $\bQ = \pi /3$ configurations. The first one
(LOFF) corresponds to a Fermi sea. A possible pair-scattering process
is indicated with arrows. We show a breached LOFF configuration as a
second example. The $\bk$ momentum values are the same as in LOFF.}
\label{fig2}
\end{figure}

Our exactly-solvable model is valid for arbitrary $\bQ$ values.
The $\bQ={\bf 0}$ limit restores the homogeneous BCS phase
giving rise to a breached-pair phase in terms of ${\bf 0}$-momentum pairs,
while a finite value of  $\bQ$ gives rise to the LOFF phase
with $\bQ$-momentum pairs.
For an asymmetric system with an excess of the $a$ species ($N_a > N_b$),
the atoms fill the lowest states up to $\bk_F^b$ with
$|\bk_F^{a}|>|\bk_F^{b}|$ at weak coupling.
When the interaction is switched on,
several possible states compete to determine the absolute GS.
The position of the unpaired atoms, defining the seniority quantum
numbers $\nu_{\bk,\bQ}$, block the available states from scattering
pairs of atoms effectively reducing the degeneracies to $d_{\bk,\bQ}$.
When $\bQ={\bf 0}$ the equations reduce to the well-known Richardson
model with blocked states \cite{RichN,NB}. In general, configurations
are identified by their $g \rightarrow 0$ limit, with specific pair and
seniority occupations. They can be categorized as follows:
\vspace*{-0.8cm}
\begin{Quote}
\item {\sf asymmetric BCS (aBCS)}:
       $\bQ={\bf 0}$, $a$ and $b$ particles fill their lowest orbitals
       up to their corresponding Fermi levels.
\item {\sf breached A}: same as {\sf aBCS},
       but the unpaired $a$ particles move up in energy
       such that pairing correlations can develop around $\bk_F^{b}$.
\item {\sf breached B}: same as {\sf aBCS},
       but the unpaired $a$ particles move down in energy
       such that pairing correlations can develop around $\bk_F^{a}$.
\item {\sf LOFF}:
       finite $\bQ$, $a$ and $b$ particles fill their lowest orbitals
       up to their corresponding Fermi levels.
\item {\sf breached LOFF}:
       finite $\bQ$, but now some of the unpaired $a$ particles move
       to allow more pairing correlations.
\end{Quote}
We illustrate some of these states by using a $D=1$ lattice with
$L=12$ modes as an example. In Fig. \ref{fig1} we show the level scheme
for a  system with $N_a=9$ and $N_b=5$. In the first column, labeled
aBCS, the excess of $a$ atoms occupy the states between $\bk_F^{b}$ and
$\bk_F^{a}$ completely blocking these states. The corresponding $b$
states are represented by a dash line. Pair scattering can only occur
between states below $\bk_F^{b}$ and states above $\bk_F^{a}$, as
indicated in the figure. It is worth noting that the excess of $a$
atoms could be located in any configuration in $\bk$-space. In the
second column we display a breached-pair superfluid state
\cite{sarma,liu,bedaque} (Breached A) where the unpaired $a$ atoms are
promoted to higher-energy states to leave some space around $\bk_F^{b}$
for pairing. A possible pair-scattering process is indicated in the
figure. Alternatively, the blocked states could be moved down for the
pair scattering to take place around $\bk_F^{a}$ as shown in the third
column of the same figure (Breached B). The relative stability of each
one of these possible states will depend upon the competition between
the kinetic energy and the pairing interaction.

Figure \ref{fig2} displays two examples of a LOFF state configuration.
Here we assume a momentum $\bQ$ that exactly matches the two Fermi
energies ($\bQ=\bk_F^{a}-\bk_F^{b}=\pi/3$). In the first example,
corresponding to the first two columns of Fig. \ref{fig2}, the atoms
occupy the lowest single-particle energies $\varepsilon_\bk^\alpha$
defining a configuration which is expected to be the lowest LOFF state
at weak coupling. The numbers within the circles indicate the {$\bQ =
\pi /3$} momentum pairs. The unpaired $a$ atoms, displayed with a black
circle, block the corresponding states of the $b$ atoms.
A possible breached LOFF configuration is shown as a second example.

We will now explore the competition between the possible phases in
a numerical example for $D=2$. We assume a square lattice with
dispersion $\varepsilon_{\bk}^\alpha=-2(\cos(k_x)+\cos(k_y))$,
with units chosen such that $k_x$ and $k_y$ are multiples of $2\pi/L$,
where $L$ is the linear size of the lattice. Being the dispersion equal
for both atomic species, we are excluding an asymmetry in the masses or
a deformed Fermi surface.  Preliminary results for asymmetric masses do
not show  qualitative differences with the results presented below,
except that for $m_a>m_b$ breached A will be more stable than breached
B because particles of type $a$ will require less kinetic energy to
shift to higher momenta than particles of type $b$.

\begin{figure}[htb]
\hspace*{-0.3cm}
\includegraphics[angle=0,width=7.5cm,scale=1.]{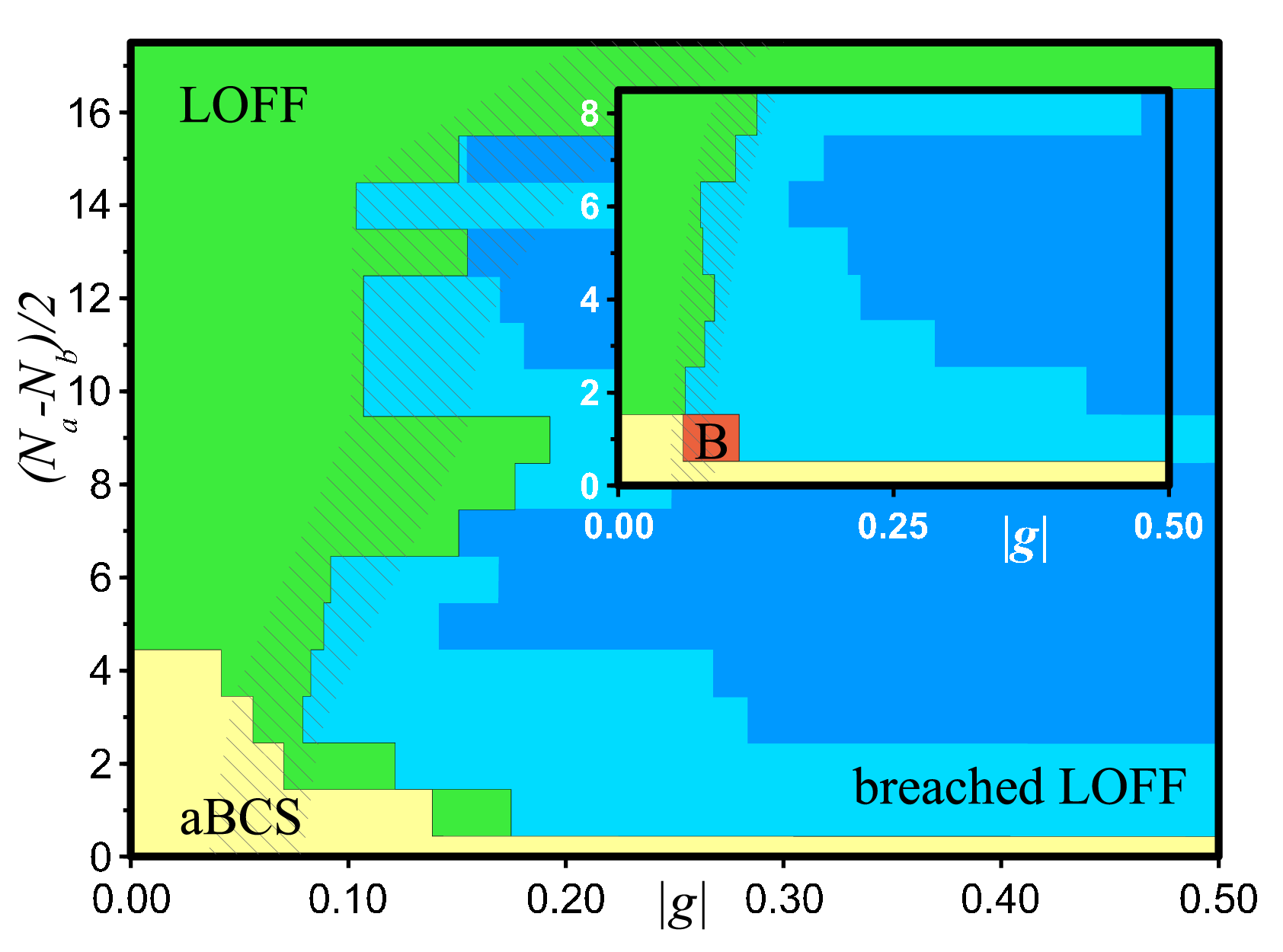}
\caption{Quantum phase diagram for a $6\times 6$ lattice at half
     filling; the inset displays the quarter-filling case (aBCS:
     yellow, LOFF: green, breached B: red, breached LOFF: blue,
     dark blue for $\bQ =(\pi,\pi)$).
     The shaded area indicates the
     transition from the normal to the superfluid phase~\cite{elec}.
\label{fig3}}
\end{figure}

\begin{figure}[htb]
\hspace*{-0.3cm}
\includegraphics[angle=0,width=7.cm,scale=1.]{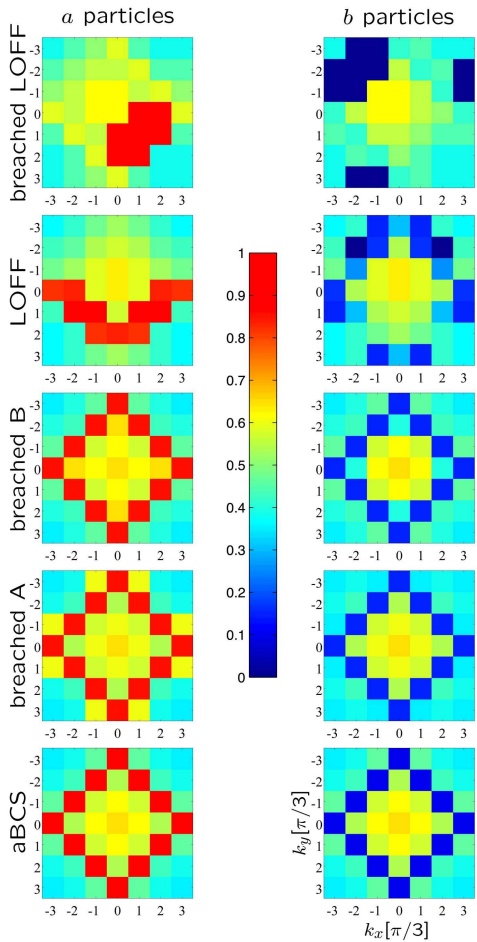}
\caption{Occupation numbers in momentum space for various
   configurations with particle numbers $N_a=22$ and $N_b=14$, at
   $g=-0.5$, for  a $6\times6$ lattice (LOFF:$\bQ =(\pi/3,0)$; breached
   LOFF:$\bQ =(\pi/3,\pi/3)$). Occupation numbers for the unpaired particles
   have been symmetrized over all possible orientations.
\label{fig4}}
\end{figure}

We discuss first the limiting cases  of a very weak or very strong
pairing interaction.  If the interaction strength $g$ is much smaller
than the level spacing,  then the full problem reduces to a pairing
problem for each level separately,  with the coupling between levels
entering at order $g^2$.  The leading order is the single-particle
energy.  This means that the GS fills the lowest single-particle
orbitals  up to the Fermi levels of species $a$ and $b$, respectively.
This leaves no room for the breached-pair phase in the GS.  aBCS and
LOFF are degenerate to leading order.  This degeneracy is lifted at
first or second order in $g$.  The pairing interaction favors open
shells,  hence it prefers a non-zero value of $\bQ$  when the valence
shell is completely filled,  while it might prefer $\bQ={\bf 0}$ for
open valence shells.  For the $6 \times 6$ model studied here,  we
found that aBCS dominates the weak limit for $N_a-N_b<4$ at quarter
filling and for $N_a-N_b<10$ at half filling.

In the strong-coupling limit one can expand Eq. (\ref{Rich}) in terms of
$g^{-1}$. In this way, the asymptotic GS of (\ref{hamil}) can be
analytically determined~\cite{Yuzba03}.
The  resulting GS energy is given by
\begin{eqnarray}
E_0 &=&  -2 g N_b (2\Omega+1 -N_b) + \sum_\bk \varepsilon_{\bk}^a  \nu_\bk^a
\nonumber  \\
&+& 2N_b \ \frac{\sum_\bk d_{\bk,\bQ} \ \eta_{(\bk,\bQ)}} {\sum_\bk
d_{\bk,\bQ} } + {\mathcal{O}(g^{-1})},
\label{eglarge}
\end{eqnarray}
with $\Omega=\sum_\bk \Omega_{\bk,\bQ}$.  Upon inspection one finds
that for the lattice model considered here,  the lowest possible value
for $E_0$ will occur when $\bQ=(\pi,\pi)$.  In that case all
$\eta_{(\bk,\bQ)}$ vanish, and the first line of Eq. (\ref{eglarge})
becomes an exact expression with the excess $a$ atoms occupying the
lowest single-particle states.  One can describe this regime as an {\it
extreme} breached LOFF state.  This result is exact in the limit of
$|g|$ much larger than the bandwidth, which is an unphysical
assumption.  However, it indicates that at some finite value of $g$  a
transition to an exotic inhomogeneous phase must occur,  combining a
{\em breached} configuration with a non-zero value for $\bQ$.  We find
such configurations to have a lower energy than the aBCS,  breached A
or B or LOFF configurations at interaction strengths as weak as
$g=-0.1$, which might be realizable in a physical setting.

We have studied model (\ref{hamil}) numerically on a $6 \times 6$
lattice,
with 18
(quarter-filled) or 36 (half-filled) particles,   distributed over
$a$ and $b$ states.   Finding the optimal configuration for the
unpaired particles turned out to be highly non-trivial because of
the large number of possibilities. We addressed this problem using
a quantum Monte Carlo technique~\cite{Romb05} that provided a
number of candidate GS configurations for various values of $g$
and $\bQ$. Starting from the $g=0$ configurations, the solution of
Eq. (\ref{Rich}) was then obtained by slowly increasing the value
of $g$, and by applying the iteration techniques explained in
Ref. \cite{Romb04}.  In this way we evaluated the exact energies
for each configuration  up to $g=-0.5$,
and we were able to determine the GS and the exact transition points.
One can see in Fig. \ref{fig3} that exotic configurations
such as LOFF or breached LOFF can have a lower energy than the aBCS state.
There is a subtle competition between LOFF and the various
breached BCS states,  and both phenomena appear simultaneously
in the emergent breached-LOFF regime that dominates the phase diagram
at larger asymmetries and interaction strengths.
The recently observed \cite{Ket} normal state region at weak coupling
could be qualitatively determined using the techniques discussed in \cite{elec}.
%
Our approach also allows computation of the occupation numbers in momentum
space.  They are derived from the integrals of motion using the
Hellman-Feynman theorem. Figure \ref{fig4} shows results for a selected
number of configurations, corresponding to the lowest-lying states of
the half-filled model at $N_a-N_b=8$, to illustrate the aBCS, breached
A and B, LOFF and breached LOFF phases at $g=-0.5$.

In summary, we presented a one-channel exactly-solvable model that
admits several homogeneous and inhomogeneous phases depending upon
the relative strength between kinetic and pairing interactions,
and the difference in the number of atomic species (given a fixed
total number of atoms).  The inhomogeneous phases (LOFF) show up
as soon as the difference in Fermi momentum between the two
species becomes commensurate with the unit lattice momentum. A
most significant result is the prediction of a new exotic phase
which combines pairs with definite momentum and breached
superfluidity/superconductivity, that we dubbed breached LOFF. We
expect this new phase to be the stable ground state at large
interaction strengths for fixed, asymmetric particle numbers.
These phases can be experimentally differentiated in
time-of-flight measurements of the molecular velocity, after
sweeping the system through the BCS-to-BEC crossover region
\cite{Regal,becbcs}. The momentum distribution of unpaired
fermions may distinguish the various exotic phases discussed here.
The present analysis can also be extended to a two-channel
integrable model with the explicit treatment of the Feshbach
resonance \cite{phase2} in a similar way as in Ref. \cite{AtMol}.

We acknowledge discussions with J. Carlson, W. V. Liu, C. Lobo, S.
Reddy and D. Van Neck.  JD, SR and KVH acknowledge financial support
from the Spanish DGI under grant BFM2003-05316-C02-02, and the FWO
- Flanders.

\end{document}